\documentstyle{article}

\def\greaterthansquiggle{\raise.3ex\hbox{$>$\kern-.75em\lower1ex\hbox{$\sim$}}}

\def\lessthansquiggle{\raise.3ex\hbox{$<$\kern-.75em\lower1ex\hbox{$\sim$}}}

\newcommand{\beq}{\begin{equation}}
\newcommand{\sgn}{\ \mbox{sgn}}

\newcommand{\eeq}{\end{equation}}
\newcommand{\beqa}{\begin{eqnarray}}
\newcommand{\eeqa}{\end{eqnarray}}
\newcommand{\beqan}{\begin{eqnarray*}}
\newcommand{\eeqan}{\end{eqnarray*}}
\newcommand{\ba}{\begin{array}}
\newcommand{\ea}{\end{array}}

\newcommand{\Ra}{\Rightarrow}
\newcommand{\ve}{\varepsilon}
\newcommand{\ep}{\epsilon}
\newcommand{\vp}{\varphi}

\newcommand{\wt}{\widetilde}
\newcommand{\wh}{\widehat}

\newcommand{\cL}{{\cal L}}

\newcommand{\dfrac}{\displaystyle \frac}

\def\nz{\ifmmode {I\hskip -3pt N} \else {\hbox {$I\hskip -3pt N$}}\fi}
\def\zz{\ifmmode {Z\hskip -4.8pt Z} \else
       {\hbox {$Z\hskip -4.8pt Z$}}\fi}
\def\qz{\ifmmode {Q\hskip -5.0pt\vrule height6.0pt depth 0pt
       \hskip 6pt} \else {\hbox
       {$Q\hskip -5.0pt\vrule height6.0pt depth 0pt\hskip 6pt$}}\fi}
\def\rz{\ifmmode {I\hskip -3pt R} \else {\hbox {$I\hskip -3pt R$}}\fi}
\def\cz{\ifmmode {C\hskip -4.8pt\vrule height5.8pt\hskip 6.3pt} \else
       {\hbox {$C\hskip -4.8pt\vrule height5.8pt\hskip 6.3pt$}}\fi}
\def\au{{\setbox0=\hbox{\lower1.36775ex%
\hbox{''}\kern-.05em}\dp0=.36775ex\hskip0pt\box0}}
\def\ao{{}\kern-.10em\hbox{``}}

\newtheorem{Theorem} {Theorem} [section] 
\newtheorem{Corollary} [Theorem] {Corollary}

\newtheorem{Proposition} [Theorem] {Proposition}

{\catcode `\@=11 \global\let\AddToReset=\@addtoreset}
\AddToReset{equation}{section}

\newcommand{\subjclass}[1]{}

\newcommand{\bysame}{---} 

\newcommand{\proof}{{\sc Proof:}\ }
\def\scri{\hbox{${\cal J}$\kern -.645em {\raise
      .57ex\hbox{$\scriptscriptstyle (\ $}}}}

\newcommand{\eq}[1]{(\ref{#1})}

\newcommand{\commentout}[1]{}

\newcommand{\ee}{\end{equation}} \newcommand{\bea}{\begin{eqnarray}}
\newcommand{\eea}{\end{eqnarray}}
\newcommand{\beaa}{\begin{eqnarray*}}
\newcommand{\eeaa}{\end{eqnarray*}} 

\newcommand{\ra}{\rightarrow}

          \newcommand{\R}{{\Bbb R}}

\newcommand{\Bbb}{\bf}
\begin{document} 

\title{Killing vectors in asymptotically flat space--times: I.
Asymptotically translational Killing vectors and the rigid positive
 energy theorem}

\author{Robert Beig\thanks{Supported by Fonds zur F\"orderung der
wissenschaftlichen Forschung, Project P9376--PHY.{\em E--mail}: 
Beig@Pap.UniVie.AC.AT}
  \\Institut f\"ur Theoretische Physik\\ Universit\"at Wien\\ A--1090
  Wien, Austria\\ \\ Piotr T.\ Chru\'sciel\thanks{ 
 On leave of absence from the Institute of
    Mathematics, Polish Academy of Sciences, Warsaw.
  Supported in part by KBN grant \# 2P30105007,
 by the Humboldt Foundation
 and by the Federal Ministry of Science and Research, Austria.
  {\em E--mail}:
    Chrusciel@Univ-Tours.fr} 
\\ D\'epartement de Math\'ematiques\\Facult\'e
des Sciences\\ Parc de Grandmont\\ F37200 Tours, France}

\maketitle
\begin{abstract} 
We study Killing vector fields in asymptotically flat
space--times. We prove
the following  result,  implicitly assumed in the uniqueness
theory of stationary black holes.
If the conditions of the rigidity part of the 
positive energy theorem are met, then in such space--times there 
are no asymptotically null Killing vector fields except if the initial
data set can be embedded in Minkowski space--time. 
We also give a  proof of the non--existence of non--singular (in an
appropriate sense) asymptotically flat space--times which satisfy an 
energy condition and
which have a null ADM four--momentum, under conditions weaker
than previously considered. 
\end{abstract}

\section{Introduction}
\label{introduction}

A prerequisite for an analysis of stationary black holes is the
understanding of properties of Killing vector fields in asymptotically
flat\footnote{Recall that there exist various papers analyzing
  properties of Killing vector fields in asymptotically flat
  space--times \cite{AAM,AshtekarSchmidt,AX,BSchmidt}. These papers do
  not, however, seem to give answers to the questions asked here.
  Moreover,  the asymptotic conditions here are
  considerably weaker than considered in those references.}
space--times. Consider, for instance, an asymptotically flat partial
Cauchy surface $\Sigma$ in a space--time $(M,g_{\mu\nu})$ with a Killing 
vector
field $X^\mu$. In the case of a stationary black hole one is interested in
situations where $X^\mu$ is timelike in the asymptotic regions. [Here we
say that an asymptotically flat space--time $(M,g_{\mu\nu})$ with a Killing
vector field $X^\mu$ is stationary if $X^\mu$ is timelike in the asymptotic
regions of $M$.] A natural question to ask is, how does then $X^\mu$
behave in the asymptotic regions? Now it is easily seen from the
equations
\begin{equation}
\label{I.0}
\nabla_\mu\nabla_\nu X_\alpha = {R^\lambda}_{\mu\nu\alpha}X_\lambda
\end{equation}
(which are a well known consequence of the Killing equations) and from
the asymptotic flatness conditions ({\em cf.\/} Propositions
\ref{2PN.1} or \ref{PD.1}, Section
\ref{KVSH}, 
for a precise
description of the asymptotic conditions needed here) that there exist
constants $A^\mu$ such that every Killing vector field $X^\mu$ which is
timelike for $r\ge R$ for some $R$ satisfies
\begin{eqnarray}
& X^\mu - A^\mu \rightarrow_{r\to\infty} 0\, ,\label{I.1} &
\\
& \eta_{\alpha\beta}A^\alpha   A^\beta \le 0 \,. &
 \nonumber
\end{eqnarray}
Here $\eta_{\alpha\beta}$ is the Minkowski metric, and we use the
signature $(-,+,+,+)$. It should be emphasized that the requirement of
timelikeness of $X^\mu$ for large $r$ does {\em not\/} exclude the
possibility that $\eta_{\alpha\beta}A^\alpha A ^\beta $ vanishes. Indeed, 
an
explicit example of a metric (not satisfying any reasonable field
equations) with an everywhere timelike Killing vector which is
asymptotically null can be found in \cite{ChWald} ({\em cf.\/} the
Remark preceding Theorem A.1, Appendix A of \cite{ChWald}). 
(Let us point out that by a null vector we mean a non--zero vector of zero 
Lorentzian length.)
Now in the
uniqueness theory of black holes it is customary to assume that $A^\mu
= \delta^\mu_0$ in an asymptotically flat coordinate system in which
$\Sigma$ is given by an equation $x^0=0$. If the orbits of the Killing
vector field $X^\mu$ are complete (at least in the asymptotic region) and
if $A^\mu$ is timelike, then  $\Sigma$ can
 be deformed (``boosted") to a new partial Cauchy surface for
which $A^\mu = \delta^\mu_0$ (in an appropriately redefined
asymptotically flat coordinate system). If, however, $X^\mu$ is
asymptotically null (by which we mean that the vector $A^\mu$
appearing in (\ref{I.1}) is null), then no such deformation is
possible and we are faced with the intriguing possibility of existence
of stationary space--times in which the Killing vector cannot be
reduced to a standard form where the metric is diagonal and the vector
$A^\mu $ of   (\ref{I.1}) equals  $\delta^\mu_0$.
As has been argued in \cite{Chnohair}, the existence of such Killing
vector fields does not seem to be compatible with the rigidity part of
the positive energy theorems. Here we make the arguments of
\cite{Chnohair} precise and show the following (the reader is referred
to Theorem \ref{TV.1} for a more precise formulation):

\begin{Theorem}
  \label{T1new}
  Let $(M,g_{\mu\nu})$ be a space--time with
a Killing vector field which is  asymptotically null along an
(appropriately regular) 
asymptotically flat spacelike hypersurface $\Sigma$.  
 Then the  ADM energy--momentum vector of $\Sigma$ vanishes.
\end{Theorem}

To say more about space--times considered in Theorem \ref{T1new}
one can use the positive energy theorem. In Section \ref{pets} below we 
prove the following\footnote{Various variants of Theorem \ref{Tpet3} are 
of course well--known, {\em cf.\/} Section \ref{pets} for a detailed 
discussion.}:

\begin{Theorem}[``Timelike ``future--pointing'' four--momentum theorem''] 
\label{Tpet3}
Under the conditions of Theorems \ref{Tpet} and \ref{Tpet2} below, suppose 
that the initial data $(\Sigma,g_{ij},K_{ij})$ are {\em not} initial
data for Minkowski space--time. Then the  ADM energy--momentum vector $p^ 
\mu
$ of $\Sigma$ satisfies
$$
p^0 > \sqrt{\sum_{i=1}^3(p^i)^2}.
$$
\end{Theorem}

Theorem \ref{T1new} can be used
together with 
Theorem \ref{Tpet3} to obtain the following:

\begin{Theorem}  \label{T1}
  Let $(M,g_{\mu\nu})$ be a maximal globally hyperbolic space--time with a 
Cauchy surface
 satisfying the requirements of Theorems \ref{Tpet} and \ref{Tpet2}. 
 Let $X^\mu$ be a Killing vector field on $M$
  which is asymptotically null along an asymptotically flat Cauchy
surface. Then $X^\mu$ is 
  everywhere null and $(M,g_{\mu\nu})$ is the Minkowski space--time.
\end{Theorem}

Theorem \ref{T1} and the results of \cite{Chorbits} ({\em cf.\/} also
\cite{Chnohair}[Theorem 1.7]) show that there is no loss of generality
in assuming that $A^\mu = \delta^\mu_0$ in, say electrovacuum,
maximal globally hyperbolic space--times with an appropriately regular
asymptotically flat Cauchy surface.
Let us mention that the results here settle in the positive Conjecture
1.8 of \cite{Chnohair}.

  This paper is organized as follows. In Section \ref{KVSH} we
discuss some general properties of Killing vector 
fields in asymptotically flat space--times. In order to minimize the
number of assumptions  we  adopt a $3+1$ dimensional
point of view; the various advantages  for doing  that are explained
at the beginning of Section \ref{KVSH}. The main result there are
Proposition \ref{2PN.1} and \ref{PD.1} which establish the asymptotic
behaviour of Killing vectors along asymptotically flat spacelike surfaces.
In that section we also introduce the notion of {\em Killing
development}, which turns out to be very useful in our analysis.
In section \ref{translational} we study the relationship between the
ADM four--momentum and the asymptotic behaviour of the Killing
vector. The results there can be summarized as follows: If
$X^\mu\to_{r\to\infty} A^\mu$ along an asymptotically flat spacelike
surface $\Sigma$, then the ADM four--momentum is  proportional
to $A^\mu$. The proportionality constant is zero when $A^\mu$ is not
timelike. 
Let us point out, that some similar  results can be found in
\cite{AAM}. However in that reference the possibility of
asymptotically null Killing vector fields is not taken into
consideration. Also, in \cite{AAM} rather strong asymptotic conditions
are imposed. In a sense most of the work here consists in showing that the
asymptotic conditions needed to be able to obtain the desired
conclusions can actually be derived from the decay conditions on the
matter sources and from the hypothesis of existence of Killing vector
fields.
In Section \ref{pets} we prove a positive energy theorem with
hypotheses and 
asymptotic conditions appropriate for our purposes. Theorems
\ref{Tpet} and \ref{Tpet2} there are improvements of known results,
{\em cf.\/} the beginning of Section \ref{pets} for a more detailed
discussion.

\section{Killing vectors and spacelike hypersurfaces}
\label{KVSH}
Consider a space--time $(M,g_{\mu\nu})$ with a Killing vector field
$X^\mu$, 
\beq
\label{K.1}
\nabla_\mu X_\nu + \nabla_\nu X_\mu =0\ ,
\eeq
where $\nabla_\mu$ is the covariant derivative operator of the metric
$g_{\mu\nu}$. Let $\Sigma$ be a spacelike hypersurface in $M$ and
suppose that on $\Sigma$ the field of unit normals $n^\mu$ can be
defined; this will be the case {\em e.g.\/} if $(M,g_{\mu\nu})$ is
time--orientable.  On $\Sigma$ define a scalar field $N$ and a vector
field $Y^i$ by the equations
\begin{eqnarray}
\label{K.2}
& N = - n_{\mu} X^\mu\ , &
\\
\label{K.3}
& g_{ij} Y^i dx^j = i^* (g_{\mu\nu}X^\mu dx^\nu)\ ,
&
\end{eqnarray}
where $i$ denotes the embedding of $\Sigma$ into $M$. We use the
symbol $g_{ij}$ to denote the pull--back metric $i^* g_{\mu\nu}$. 
Eq.\ \eq{K.1} with $\mu=i$ and $\nu=j$ reads
\beq
\label{K.4}
2NK_{ij} + {\cal L}_Y g_{ij} = 0 \ ,
\eeq
where $\cal L$  denotes the Lie derivative, and $K_{ij}$ is the
extrinsic curvature tensor of $i(\Sigma)$ in $(M,g_{\mu\nu})$, defined
as\footnote{$K_{ij}$ as defined here is $-K_{ij}$ as 
in \cite{YorkinSmarr}; similarly $J^i$ as defined here is $-J^i$ as
defined there.\label{extrinsiccurvature}}
the pull--back to $\Sigma$ of $\nabla_\mu n_\nu$. 
Set
$$
\Sigma_{N>0} = \{p\in\Sigma : N\ne 0\} \ .
$$
In a neighbourhood of $\Sigma_{N>0}$ we can introduce a coordinate
system $(u,x^i)$ in which $X^\mu\partial_\mu = \partial _u$ and in
which $\Sigma_{N>0}$ is given by the equation $u=0$. The metric on
this neighbourhood takes the form
\beq
\label{K.0}
g_{\mu\nu}dx^\mu dx^\nu = -N^2 du^2 + g_{ij}(dx^i+Y^idu)(dx^j+Y^jdu)\
, 
\eeq
with some functions which do not depend upon $u$.
Let $G_{\mu\nu}$ be the Einstein  tensor of $g_{\mu\nu}$, that
is,  $G_{\mu\nu}= R_{\mu\nu}-\frac{g^{\alpha\beta}R_{\alpha\beta}}{2}
g_{\mu\nu}$, where $R_{\mu\nu}$ is the Ricci tensor of $
g_{\mu\nu}$. We have the $3+1$ decomposition formulae ({\em cf.\
e.g.\/} \cite{YorkinSmarr})
\begin{eqnarray}
& 2G_{\mu\nu} n^\mu n^\nu={}^3R+K^2-K^{ij}K_{ij}\ ,
\label{K.5}  
& \\ &
G_{i\mu}n^\mu = D_j(K^{ij}-g^{kl}K_{kl}g^{ij}) \ , \label{K.6}
& \\ &
G_{ij}-\frac{1}{2}g^{k\ell}G_{k\ell}g_{ij} =
{}^3R_{ij} + K K_{ij} - 2 K_{ik} K^k{}_j
\qquad\qquad\qquad\qquad\qquad \nonumber
& \\ &
\qquad \qquad \qquad\qquad\qquad - N^{-1}(\cL_Y K_{ij} + D_i D_j N  )
-\frac{1}{2}G_{\mu\nu} n^\mu n^\nu\,g_{ij} \ .
\label{K.7}
\end{eqnarray}
Here $g^{ij}$ is the tensor inverse to $g_{ij}$, $K =g^{kl}K_{kl}$, 
${}^3R_{ij}$ is the
Ricci tensor of the metric 
$g_{ij}$, and ${}^3R=g^{ij}{}^3R_{ij}$.
All the above  is of course well--known, we have written it down in
detail to fix the notation and to spell--out the conditions needed for
the definition of the fields $N$ and $Y^i$. In particular we wish to
emphasize that we did not need to assume completeness of the orbits of
$X^\mu$, we did not need to assume that the orbits of $X^\mu$ intersect
$\Sigma$ only once, etc. It is however the case that those last
properties are needed for several arguments, {\em e.g.\/} in the
uniqueness theory of black holes ({\em cf.\ e.g.\/}
\cite{Chnohair}). By way of  example, consider a maximal globally 
hyperbolic
space--time $(M,g_{\mu\nu})$ with an asymptotically flat Cauchy
surface with compact interior, with a metric satisfying the
Einstein--Yang-Mills--Higgs equations, and with a Killing vector field
$X^\mu$. While one expects the orbits of $X^\mu$ to be complete ({\em
cf.\ e.g.\/} \cite{Chorbits} for an analysis in the vacuum case), no
proof of such a result has been established so far. It is therefore of
interest  to establish various properties of space--times 
 $(M,g_{\mu\nu})$ with Killing vectors with a minimal amount of global
assumptions on $M$. As one is often interested in globally hyperbolic
space--times it does not seem to be overly restrictive to assume the
existence in $M$ of a spacelike hypersurface $\Sigma$ satisfying the
hypotheses spelled out at the beginning of this section. The
construction above yields then a scalar field $N$ and a vector field
$Y^i$ defined on $\Sigma$, such that eqs.\ \eq{K.4}--\eq{K.7}
hold. Consider then a set $(\Sigma,g_{ij},K_{ij},N,Y^i)$. We shall
call the {\em Killing development of $(\Sigma,g_{ij},K_{ij},N,Y^i)$}
the space--time $(\hat M, \hat g_{\mu\nu})$, where $$\hat M =
{\R}\times \Sigma_{N>0}\ ,$$
and where $\hat g_{\mu\nu}$ is given by the equation
\begin{eqnarray}
\label{K.10} &
\hat g_{\mu\nu}dx^\mu dx^\nu = -\hat N^2 du^2 + \hat g_{ij}(dx^i+\hat
Y^idu)(dx^j+\hat Y^jdu)\ , 
& \\ \nonumber
& 
\hat N(u,x^i) = N(x^i),\quad
\hat g_{ij}(u,x^i) = g_{ij}(x^i),\quad
\hat Y^i(u,x^i) = Y^i(x^i)\ . &
\end{eqnarray}
Here the $u$ coordinate runs over the $\R$ factor in $
{\R}\times \Sigma_{N>0}$. Clearly the vector field
$X^\mu\partial_\mu=\partial_u$ is a Killing vector, so that
\beq
\label{K.10.0}
\hat\nabla_\mu X_\nu + \hat\nabla_\nu X_\mu =0\ ,
\eeq
  where $\hat\nabla_\mu$ is the covariant derivative operator of the metric
$\hat g_{\mu\nu}$. Note that
\beq
\label{K.11}
X_i\Big|_{u=0} = Y_i,\qquad \hat N\Big|_{u=0} = N\ .
\eeq
Consider the extrinsic curvature tensor $\hat K_{ij}$ of the slices
$u=\mbox{const}$. In general $\hat K_{ij}$ will have nothing to do
with the tensor field $K_{ij}$. Suppose, however, that \eq{K.4}
holds. Eq.\ \eq{K.10.0} with $i=\mu$ and $\nu=j$, eq.\ \eq{K.11} and
\eq{K.4} give then, at $u=0$,
\beq
\label{K.12}
\hat K_{ij}=K_{ij}\ .
\eeq
Since $\hat K_{ij}$ is $u$--independent it follows that this last
relation holds throughout $\hat M$. One also notices that \eq{K.12}
will hold if and only if \eq{K.4} holds.

Consider, next, the Einstein  tensor   $\hat G_{\mu\nu}$ of
the metric $\hat g_{\mu\nu}$. It is given by the hatted equivalent of
eqs.\ \eq{K.5}--\eq{K.7}. Given the set 
$(\Sigma,g_{ij},K_{ij},N,Y^i)$ one can define on $\Sigma_{N>0}$ a
scalar field $\rho$, a vector field $J^i$, and a tensor field
$\tau_{ij} $ via the equations
\begin{eqnarray}
\label{K.13}
&
 2 \rho = {}^3R+K^2-K^{ij}K_{ij}\ , 
& \\ &
J^i=D_j(K^{ij}-Kg^{ij}) \ , 
& \label{K.14} \\ &
\tau_{ij}-\frac{1}{2}g^{k\ell}\tau_{k\ell}g_{ij} =
{}^3R_{ij} + K K_{ij} - 2 K_{ik} K^k{}_j
\qquad\qquad\qquad\qquad \nonumber
& \\ &
\qquad \qquad \qquad\qquad\qquad 
- N^{-1}(\cL_Y K_{ij} + D_i D_j N  )
-\frac{\rho}{2}\,g_{ij} \ . &
\label{K.15}
\end{eqnarray}
If eq.\ \eq{K.4} holds it follows from \eq{K.11}--\eq{K.12} that we
will have
\beq
\label{K.16}
\hat G_{\mu\nu} \hat n^\mu \hat n^\nu (u,x^\ell)= \rho (x^\ell) , \quad
\hat G_{i\nu} \hat n^\nu (u,x^\ell) = J_i (x^\ell),  \quad
\hat G_{ij} (u,x^\ell) = \tau_{ij} (x^\ell)\ ,
\eeq
where $\hat n_\mu$ is the unit normal to the slices $u =$ const.
It is of interest to consider the case of covariantly constant Killing
vector fields. In that case on a hypersurface $\Sigma$ as at the
beginning of this 
section we will have
\begin{eqnarray}
\label{K.01}
& N K_{ij}+ D_iY_j = 0 \ , &
\\
\label{K.02}
&  K_{ij}Y^j+ D_i N = 0 \ . &
\end{eqnarray}
Let us show that if \eq{K.01}--\eq{K.02} hold, then the vector field
$X^\mu\partial_\mu=\partial_u$ on the Killing development $(\hat
M,\hat g_{\mu\nu})$ of  
$(\Sigma,g_{ij},K_{ij},N,Y^i)$ will be covariantly constant. To see
that note that eqs.\ \eq{K.01}, \eq{K.11} and \eq{K.12}  imply
$$
\hat \nabla_i X_j = 0
$$
at $u$=0, hence throughout $\hat M$. Eq.\ \eq{K.02} similarly gives
$$
\hat \nabla_i X_0 = 0\ .
$$
As $X^\mu$ satisfies \eq{K.10.0} the equations $
\hat \nabla_\mu X_\nu = 0
$ readily follow.

In our work, as well as in various other analyses, an essential role is
 played by the asymptotic
behaviour of the Killing vector fields. 
Let us start with a result based on a four--dimensional formalism.
 For $R> 0$ let $M_R$ be defined by
\begin{equation}
M_R=\big\{(t,\vec x)\in {\R} \times \big({\R}^3\setminus
B(R)\big)\big\}\,, 
  \label{D.0}
\end{equation}
where $B(R)$ is a closed ball of radius $R$. Let $\alpha$ be a
positive constant; the couple $(M_R,g_{\mu\nu})$ will be called an {\em
$\alpha$--asymptotically flat four--end} if the Lorentzian metric $g$
defined on $M_R$ is twice differentiable\footnote{In this paper for 
several 
  purposes we could assume weak differentiability of $g$ only, and replace
  the decay conditions \eq{D.1}--\eq{D.3} by some weighted Sobolev
  conditions. For the sake of simplicity we shall, however, not
  consider those weaker conditions.} and if there exists a constant $C$
such that the following inequalities hold in $M_R$:
\begin{eqnarray}
  & |g_{\mu\nu}| + |g^{\mu\nu}| + r^\alpha |g_{\mu\nu}-\eta_{\mu\nu}|
  + r^{\alpha+1} |\partial_\sigma g_{\mu\nu}| +
  r^{\alpha+2} |\partial_\sigma\partial_\rho g_{\mu\nu}| \le C \,, &
\label{D.1}\\
& g_{00}\le -C^{-1},\qquad g^{00}\le -C^{-1} \,,&
\label{D.2}\\
& \forall X^i\in {\R}^3 \quad g_{ij}X^iX^j\ge C^{-1}\sum(X^i)^2 \,.&
\label{D.3}
\end{eqnarray}
Here and throughout $\eta_{\mu\nu}$ is the Minkowski metric, while
$r=\sqrt{ x^2+y^2+z^2}$. The proof of Proposition \ref{PD.1}
that follows is based on the analysis of the equations
\begin{equation}
\label{I.0.new}
\nabla_\mu\nabla_\nu X_\alpha = {R^\lambda}_{\mu\nu\alpha}X_\lambda \ ,
\end{equation}
which are a well known consequence of the Killing equations. The
arguments follow closely those of the proof of Proposition \ref{2PN.1}
below, to be found in Appendix \ref{Aproof}, and will be
omitted.

\begin{Proposition}
  \label{PD.1}
  Let $R>0$ and let $X^\mu$ be a Killing vector field defined on an
  $\alpha$--asymptotically flat end $M_R$, $0<\alpha<1$.
  Then there exist numbers $ \Lambda_{\mu\nu}= \Lambda_{[\mu\nu]} $
  and a function $C(t)$ such that on every slice $t=\mbox{const}$ we have 
\begin{equation}
|X^\mu-{\Lambda^\mu}_\nu x^\nu| + r |\partial_\sigma X^\mu -
{\Lambda^\mu}_\sigma| + r^2 |\partial_\sigma \partial_\rho X^\mu | \le
C(t) r^{1-\alpha}\,,
  \label{D.4}
\end{equation}
with ${\Lambda^\mu}_\nu\equiv \eta^{\mu\alpha}\Lambda_{\alpha\nu}$.
If ${\Lambda_{\mu\nu}}=0$, then there exist numbers $A^\mu$ and a
constant $C$ such that on $M_R$ we have
\begin{equation}
  |X^\mu-A^\mu | + r |\partial_\sigma X^\mu | + r^2 |\partial_\sigma
  \partial_\rho X^\mu | \le C r^{-\alpha}\,.
  \label{D.5}
\end{equation}
If ${\Lambda_{\mu\nu}}=A^\mu=0$, then  $X^\mu \equiv 0$.
\end{Proposition}

{\bf Remark:} Obvious  analogs   of the results of Proposition
\ref{2PN.1} below with $k>2$ hold if higher asymptotic regularity of
the metric is assumed in 
Proposition \ref{PD.1}. It also follows from Proposition \ref{2PN.1}
below that if the constant $C$ in \eq{D.1}--\eq{D.3} is replaced by a
function of $t$,
then the conclusions of Proposition \ref{PD.1} will still hold with
the constant $C$ in \eq{D.5} replaced by some function $C'(t)$. 

Our next result is the $3+1$ equivalent of Proposition \ref{PD.1}. The
reader may wish to note the following: in the $4$--dimensional
formulation the fall--off conditions on the metric ensure that the
space--time Riemann tensor vanishes at an appropriate rate. In the
$3+1$ formulation the fall--off conditions on $g_{ij}$ and $K_{ij}$
are not sufficient to guarantee that, they must be supplemented by a
fall--off condition on $\rho$ and $\tau_{ij}$. Thus the eq.\ \eq{K.100} 
below is a rather weak
equivalent of the decay conditions $R_{\mu\nu\rho\sigma} = 
O(r^{-2-\alpha})$.
The following is a  straightforward consequence of eqs.\
\eq{K.13} and \eq{K.15}
({\em cf.\/} also \cite[Theorem 3.3 and Proposition
3.2]{ChrOM}). The notation $O_k$ is defined in Appendix \ref{definitions}. 
An outline of the
proof is given in Appendix
\ref{Aproof}. 

\begin{Proposition}\label{2PN.1} Let $R > 0$ and let $(g_{ij},K_{ij})$ be 
initial data on $\Sigma_R \equiv {\Bbb R}^3 \setminus B(R)$ satisfying
\beq
g_{ij} - \delta_{ij} = O_{k} (r^{-\alpha}), \qquad
 K_{ij} = O_{k-1} (r^{-1-\alpha}), 
\label{K.99}
\eeq
with some $
k>1$ and some $0<\alpha< 1$. 
Let $N$ be a $C^2$ scalar field and $Y^i$  a $C^2$ vector field on
$\Sigma_R$ such that eqs.\ \eq{K.4}, \eq{K.13} and \eq{K.15} hold
with some $\rho$ and $\tau_{ij}$ satisfying
\begin{equation}
|\rho| + |\tau_{ij}| \le C (1+r)^{-2-\alpha}\ .
\label{K.100}
\end{equation}
  Then there exists numbers $ \Lambda_{\mu\nu}= \Lambda_{[\mu\nu]} $
  such that we have
\begin{equation}
Y^i-\Lambda_{ij} x^j = O_{k}(r^{1-\alpha}), \qquad
N+\Lambda_{0i}x^i = O_{k}(r^{1-\alpha}) \ .
\label{D.4.0}
\end{equation}
If ${\Lambda_{\mu\nu}}=0$, then there exist numbers $A^\mu$ 
 such that we have
\begin{equation}
  Y^i-A^i  = O_{k}(r^{-\alpha}), \qquad
N-A^0 = O_{k}(r^{-\alpha}) \ .
  \label{D.4.1}
\end{equation}
If ${\Lambda_{\mu\nu}}=A^\mu=0$, then  $Y^i\equiv N \equiv 0$.
\end{Proposition}

Let us remark that if $\alpha=1$, then Proposition \ref{2PN.1} holds
with the function $r^{1-\alpha}$ in the right--hand--side of eq.\ 
\eq{D.4.0} replaced by $1+|\log r|$; similarly in  \eq{D.4.1}
$r^{-\alpha}$ has to be replaced by $r^{-1}(1+|\log r|)$.

A Killing vector field for which $\Lambda_{\mu\nu}=0$ will be called
{\em asymptotically translational}. 
 
For further use let us mention the following: Consider
$(g_{ij},K_{ij})$ such that \eq{K.99} holds, and suppose that
$(N,Y^i)$ satisfy \eq{D.4.1} with some $A^0\ne 0$. 
Suppose finally that \eq{K.4} is weakened to 
\beq
\label{K.4.1}
2NK_{ij} + {\cal L}_Y g_{ij} = O_{k-1}(r^{-\beta}) \ ,
\eeq
with some $\beta \ge 1$. 
In that case \eq{K.16}   will be replaced by 
\begin{eqnarray}
& \nonumber
\hat G_{\mu\nu} \hat n^\mu \hat n^\nu - \rho = 
O_{k-1}(r^{-\min(1+\alpha,\beta)-\beta}), \quad
\hat G_{i\nu} \hat n^\nu - J_i = O_{k-2}(r^{-\beta-1}), 
& \\ &
\hat G_{ij} - \tau_{ij}= O_{k-2}(r^{-\beta-1})\ . 
& \label{K.16.1}
\end{eqnarray}

\section{ADM four--momentum in space--times with asymptotically 
translational
Killing vectors}
\label{translational}
In this section we prove the following results: Consider an
asymptotically flat space--time with an asymptotically  translational
Killing vector field $X^\mu$, that is, there exist constants $A^\mu$
such that $X^\mu\to_{r\to\infty}A^\mu$. Then:
\begin{enumerate}
\item
If $A^\mu A_\mu \ge 0$, then the ADM four--momentum $p^\mu$ vanishes.
\item
If 
$A^\mu A_\mu < 0$, then $p^\mu$ is proportional to $A^\mu$.
\end{enumerate}
We shall establish those results in the $3$ dimensional framework
discussed in Section \ref{KVSH}. Proposition \ref{2PN.1} in that
section justifies our fall--off conditions on the fields $N$ and
$Y^i$. The results here are actually slightly more general than stated
above, in that we allow for fields which satisfy the relevant Killing
equations up to terms which decay at an appropriate rate, {\em cf.\/}
below for the precise conditions.

\begin{Proposition}\label{PN.1}
 Let $R > 0$ and let $(g_{ij},K_{ij})$ be 
initial data on $\Sigma_R \equiv {\Bbb R}^3 \setminus B(R)$ satisfying
\begin{eqnarray} \label{F0.1} &
g_{ij} - \delta_{ij} = O_{2} (r^{-\alpha}), \qquad
 K_{ij} = O_{1} (r^{-1-\alpha}), 
 \qquad \alpha > 1/2, & \\
& \label{F0.2}
 J^i = O (r^{-3-\epsilon}), \qquad \rho = O (r^{-3-\epsilon}),
\qquad \epsilon > 0\ . &
\end{eqnarray}
Let $N$ be a $ C^1$ scalar field and $Y^i$  a $C^1$ vector 
field on $\Sigma_R$ such that
\beq \label{F0.1.1}
N-A^0 = O_{1} (r^{-\alpha}), \qquad
Y^i\to _{r\to\infty} A^i \ ,
\eeq 
 for some set of constants $(A^\mu) \not\equiv 0$. 
Suppose further that
\begin{equation}
2 N  K_{ij} + \cL_Y g_{ij} =  O_1 (r^{-2-\epsilon}). \label{(PN.1.0)}
\end{equation}
Let $p^\mu$ be the ADM four--momentum  of\/  $\Sigma_R$. Then:
\begin{enumerate}
\item If $A^0 = 0$, then $p^0 = 0$.
\item  If $A^0 \ne 0$, then  $p^\mu$ is proportional to 
$A^\mu$.
\end{enumerate}
\end{Proposition}

\paragraph{{\bf Remark:}} The pointwise decay estimates assumed above
can be weakened  
to weighted Sobolev spaces conditions. To avoid a tedious discussion of
technicalities  
we shall, however, not consider such fields here. 
 
\paragraph{Proof:} Without loss of generality we may assume that both 
$\alpha$ and $\epsilon$ are strictly smaller than $1$. Eq.\ \eq{F0.1.1} 
and a simple 
analysis of eq. \eq{(PN.1.0)}  ({\em cf.\
e.g.\/} the proof of Prop.\ \ref{2PN.1}, Appendix \ref{Aproof}) show 
that
\beq \label{F0.1.11}
Y^i-A^i = O_{2} (r^{-\alpha})\ .
\eeq 
 By our
asymptotic conditions eq. \eq{(PN.1.0)} can be rewritten as
\begin{equation}
g_{ij,k} A^k + Y^i{}_{,j} + Y^j{}_{,i} = - 2 A^0 K_{ij} + 
O_1(r^{-2-\epsilon})\ , \label{(PN.1.1)}
\end{equation}
 and 
 we have redefined $\epsilon$ to be $\min (\epsilon,2\alpha -1) > 0$.
The momentum--constraint equation reads
\begin{equation}
\partial_i K_{ij} = \partial_j K + O(r^{-3-\epsilon}), \label{(PN.1.2)}
\end{equation}
where $K=g^{ij}K_{ij}$.
Taking the divergence of \eq{(PN.1.1)} and using \eq{(PN.1.2)} gives
\begin{equation}
g_{ij,kj} A^k + \Delta_\delta Y^i + \partial_i (Y^j{}_{,j}) =
- 2A^0 K_{,i} + O(r^{-3-\epsilon}). \label{(PN.1.3)}
\end{equation}
Here $\Delta_\delta = \sum_i \partial_i \partial_i$. Contracting $i$ 
with $j$ in \eq{(PN.1.1)} allows us to eliminate $\partial_j Y^j$ in 
 \eq{(PN.1.3)} 
in terms of $K_{,i}$ so that \eq{(PN.1.3)} leads to
$$
\Delta_\delta Y^i = - A^0 K_{,i} - (g_{ij,j} - \frac{1}{2}
g_{jj,i})_{,k} A^k + O(r^{-3-\epsilon}).  $$ In what follows we shall
freely make use of properties of harmonic functions on $\Sigma_R$
which were established in {\em e.g.\/} \cite{Meyers,Bartnikmass,ChAFT}.
Increasing $R$ if necessary we may choose harmonic\footnote{There
arises a slight difficulty here, related to the fact that the metric
might not satisfy the conditions  \eq{F0.1} in harmonic coordinates
due to a loss of classical differentiability. In our proof  we
have  ignored that issue, assuming {\em e.g.\/} that eq.\ \eq{(PN.1.1)}
still holds in harmonic coordinates. The problem is easily cured by
keeping track of weighted--Sobolev differentiability of various error
terms which arise in our equations, making use of the estimates of
\cite{Bartnikmass}. In doing that one can verify that the 
statement of our result is  correct as stated. All the details of the
proof as written  here can be justified if a H\"older differentiability
index $\lambda$ is added in  eqs.\ \eq{F0.1}--\eq{F0.2}. In order to make 
the
argument more transparent we have chosen  to present our proof without
the introduction of weighted Sobolev spaces.}
coordinates on $\Sigma_R$, $$
\partial_i (g^{ij} \sqrt{\det g}) = 0,
$$
with 
$$
g_{ij} - \delta_{ij} = O_1(r^{-\alpha}).
$$
If $A^0 = 0$ define $\vp$ to be identically zero, otherwise let $\vp = 
O_1(r^{1-\alpha})$ be a solution of
\beq \label{PN.1.3.0}
\Delta_\delta \vp = - A^0 K.
\eeq
Setting $Z^i = Y^i - A^i - \vp^{,i}$ one is led to
$$
\Delta_\delta Z^i = O(r^{-3-\epsilon}),
$$
so that there exist numbers $\alpha^i \in {\Bbb R}$ such that
$$
Z^i = \frac{\alpha^i}{r} + O_1(r^{-1-\epsilon}).
$$
A contraction over $i$ and $j$ in \eq{(PN.1.1)} gives
\begin{equation}
Z^i{}_{,i} = - \frac{\alpha^i x^i}{r^3}  + O(r^{-2-\epsilon})
= - \frac{1}{2} g_{ii,k} A^k  + O(r^{-2-\epsilon}).
\label{(PN.1.4)}
\end{equation}
The scalar constraint equation in harmonic coordinates gives
\begin{equation}
\Delta_\delta g_{ii} = O(r^{-3-\epsilon}) \Ra g_{ii} = 3+ \frac{\beta}{r} +
O_1(r^{-1-\epsilon}), \label{(PN.1.5)}
\end{equation}
for some constant $\beta$. Eq. \eq{(PN.1.5)}
 inserted in the formula for the ADM mass
yields
\beq \label{ADMmass}
m = \dfrac{1}{16 \pi} \int_{S_\infty} (g_{ij,j} - g_{jj,i})\, dS_i
 = - \dfrac{1}{32\pi} \int g_{jj,i}\,  dS_i =  \dfrac{\beta}{8}.
\eeq
Inserting this in \eq{(PN.1.4)} one is led to
$$
\alpha^i = - 4m A^i,
$$
so that one finally obtains
\begin{equation}
Y^i = A^i\left(1 - \frac{4m}{r}\right) + \vp_{,i} + O_1(r^{-1-\epsilon}).
\label{(PN.1.6)}
\end{equation}
Suppose first that $A^0 = 0$. In this case we necessarily have $A^i 
\not\equiv 0$, and, rescaling $X^\mu \partial_\mu$ if necessary, we can
choose coordinates so that $A^i = \delta^i_z$. Eq. \eq{(PN.1.1)} now reads
\begin{equation}
g_{AB,z} = O(r^{-2-\epsilon}), \label{(PN.1.7)}
\end{equation}
\begin{equation}
(g_{zz} + 2Y^z)_{,z} = O(r^{-2-\epsilon}), \label{(PN.1.8)}
\end{equation}
\begin{equation}
g_{zA,z} = \left(\frac{4m}{r}\right)_{,A} + O(r^{-2-\epsilon}).
\label{(PN.1.9)}
\end{equation}
Let $\rho^2=x^2+y^2$. For $\rho\ge R$ eq.\ \eq{(PN.1.9)} gives
\begin{eqnarray*}
0 & = & x^A\int_{-\infty}^\infty g_{zA,z}dz \\
 & = & -4m \int_{-\infty}^\infty \frac{dz}{(1+z^2)^{3/2}} +
\int_{-\infty}^\infty O(r^{-2-\epsilon})dz\ .
\end{eqnarray*}
To estimate the second integral it is convenient to consider
separately the integrals $\int_{-\infty}^{-\rho}$, $\int_{-\rho}^\rho$ and
$\int_{\rho}^\infty$. Elementary estimates then show that this
integral is $O(\rho^{-\epsilon})$; passing to the limit
 $\rho\to\infty$ one subsequently obtains $m=0$, which 
 establishes point 1. To establish point 2, suppose that $A^0 \neq 0$. 
After a rescaling of
$X^\mu$ if necessary we can without loss of generality assume that
$A^0 = 1$. Eq. \eq{(PN.1.1)} gives thus
\begin{equation}
\ba{rcl}
K_{ij} &=& - \dfrac{1}{2} \{ Y^i{}_{,j} + Y^j{}_{,i} + g_{ij,k} A^k\}
+ O_1(r^{-1-2\alpha}) \\[7pt]
&=& - \dfrac{1}{2} \{ Z^i{}_{,j} + Z^j{}_{,i} + 2\vp_{,ij} +
g_{ij,k} A^k\} + O_1(r^{-1-2\alpha}). 
\ea \label{(PN.1.15)}
\end{equation}  
Consider the ADM momentum\footnote{The unusual sign in eq.\ \eq{(PN.1.13)} 
is
due to our convention on $K_{ij}$, {\em cf.\/} footnote
\ref{extrinsiccurvature}.} $p_i$:
\begin{equation}
p_i = - \frac{1}{8\pi} \int_{S_\infty} (K_i{}^j - K \delta_i{}^j)
dS_j, \label{(PN.1.13)}
\end{equation} 
After insertion of \eq{(PN.1.15)} in \eq{(PN.1.13)} one finds
\begin{equation}
p_i = \frac{1}{16\pi} \int_{S_\infty} (Z^i{}_{,j} + Z^j{}_{,i} + A_j
g_{ik,k}) dS_j. \label{(PN.1.16)}
\end{equation}
Here the $\varphi$ contribution drops out because of the following
calculation:
\beq
\ba{rcl}
&  \int_{S_\infty} (\Delta_\delta \varphi
\delta_{ij} - \partial_i \partial_j \varphi)dS_j 
= &\\[7pt]
& \int_{S_\infty} (\partial_k \varphi \delta_{ij} -
\partial_j \varphi \delta_{ki})_{,k} dS_j 
=& 0. 
\ea\label{dropout}
\eeq
We have also used the identities $$ g_{ij,k} A^k = (g_{ij} A^k - g_{ik}
A^j)_{,k} + g_{ik,k} A^j, $$ 
and integration by parts to rearrange the $g_{ij,k} A^k$ terms.
Inserting \eq{(PN.1.6)} in
\eq{(PN.1.16)} and using the harmonic coordinates condition one obtains
$$
p_i = m \; A_i,
$$
which had to be established. \hfill\ $\Box$

Point 1 of Proposition \ref{PN.1} suggests strongly that the ADM
four--momentum must vanish when $A^\mu$ is spacelike.
We can show that if we assume some further asymptotic
conditions on the fields under consideration. A similar
 result has been established previously  in \cite{AAM} under
rather stronger asymptotic and global conditions. 

\begin{Proposition}\label{PN.2}
Under the hypotheses of Proposition \ref{PN.1}, suppose further that
$N$ is $C^2$ and that 
\beq
N \tau_{ij} =  O (r^{-3-\epsilon})
\ .  \label{F0.1.12} 
\eeq
If 
\beq \label{F0.1.21}
(A^0)^2 < \sum_i A^iA^i\ ,
\eeq 
then $p^\mu$ vanishes. 
\end{Proposition}

\paragraph{Proof:}
It follows from eqs.\ \eq{F0.1.1}, \eq{(PN.1.0)} and \eq{F0.1.12} that
\beq \label{F0.1.111} 
Y^i- A^i =  O_{2} (r^{-\alpha}), \quad
N-A^0 = O_{2} (r^{-\alpha})\ .
\eeq 
Consider first the case $A^0=0$; by Proposition \ref{PN.1} we have
$p^0=0$.
Let $\psi$ be any function on $\Sigma_R$ such that $\psi_{,z} = N $.
Eq.\ \eq{F0.1.12} gives
$$
(K_{ij} - \partial_i \partial_j \psi)_{,z} = O(r^{-3-\epsilon}),
$$
so that by $z$--integration one obtains
$$
K_{ij} - \partial_i \partial_j \psi = O(r^{-2-\epsilon}).
$$
Inserting this in eq. \eq{(PN.1.13)} one obtains
\beq
\ba{rcl}
p_i &=& - \dfrac{1}{8\pi} \int_{S_\infty} (\Delta_\delta \psi
\delta_{ij} - \partial_i \partial_j \psi)dS_j 
\\[7pt]
&=& - \dfrac{1}{8\pi} \int_{S_\infty} (\partial_k \psi \delta_{ij} -
\partial_j \psi \delta_{ki})_{,k} dS_j 
\\[7pt]
&=& 0. 
\ea\label{dropout1}
\eeq

Consider, next, the case $A^0\ne 0$, let $(\hat M,\hat g_{\mu\nu})$ be the
 Killing development of $(\Sigma_R,g_{ij},K_{ij},N,Y^i)$ as constructed
in Section \ref{KVSH}.
As discussed in the paragraph preceding eq.\ \eq{K.16.1},
eqs.\ \eq{F0.2} and  \eq{F0.1.12} imply that the
Einstein
tensor $\hat G_{\mu\nu}$ of $\hat g_{\mu\nu}$ will
satisfy the fall-off condition 
\beq
\label{hatTfalloff}
\hat G_{\mu\nu}=O(r^{-3-\epsilon})\ .
\end{equation}
 Let $\Lambda^\mu{}_\nu$ be the matrix of 
a Lorentz transformation such that $\Lambda^0{}_\nu A^\nu = 0$. 
Let further $\Lambda\Sigma$ be the image
 under $\Lambda^\mu{}_\nu$ of
$\Sigma_R\cap \hat M$ in $\hat M$. On $\Lambda \Sigma$ the Killing vector 
$X^\mu$ satisfies $X^0\to_{r\to\infty}0$. Eq.\ \eq{hatTfalloff} shows
that we can apply the previous analysis  to conclude that the
ADM four--momentum of $\Lambda\Sigma$ vanishes. Moreover the decay
condition \eq{hatTfalloff} ensures ({\em cf.\ e.g.\/} \cite{Chremark})
that $p^\mu$ transforms as a Lorentz
vector under  Lorentz transformations of hypersurfaces, so that
the ADM four--momentum of $\Sigma_R$ vanishes as well.
\hfill$ \Box$ 

It is of interest to consider Killing vector fields which are  covariantly 
constant. As discussed in  Section
\ref{KVSH}, in such a case eqs.\ \eq{X.0.1}--\eq{X.0.2} below  will
hold  ( with $0$ on
the right--hand--sides). We have the following result, which 
does not cover asymptotically null Killing vectors: 

\begin{Proposition}\label{PN.1.1} Under the hypotheses of Proposition
\ref{PN.1}, assume moreover that $N$ is $C^2$, that
eq.\ \eq{F0.1.12} holds and that
\begin{eqnarray}  \label{X.0.1} &
 N K_{ij}+ D_iY_j = O_1(r^{-2-\epsilon}) \ , & 
\\ \label{X.0.2}
&  K_{ij}Y^j+ D_i N = O_1(r^{-2-\epsilon})  \ ,  &
\\  &
A^\mu A_\mu \neq 0.  &\nonumber 
\end{eqnarray}
Then the ADM four momentum $p^\mu$ vanishes.
\end{Proposition}

\paragraph{Proof:}  Let $(\hat M,\hat g_{\mu\nu})$ be the
Killing development of $(\Sigma_R,g_{ij},K_{ij},N,Y^i)$ as constructed
in Section \ref{KVSH}. From what is said in that
section ({\em cf.\/} the discussion following eqs.\
\eq{K.01}--\eq{K.02})  it follows that $X^\mu\partial_\mu =
\partial_u$ will satisfy 
\beq
\hat \nabla_\mu X_\nu =  O_1(r^{-2-\epsilon})\ .   \label{X.0}
\eeq
As is well known \cite{Beig,AAM}, we have
\beq
p_\mu A^\mu = \lim_{r \ra \infty} \frac{1}{8\pi} \int \hat\nabla^{[\mu}
X^{\nu]} dS_{\mu\nu} \label{X.2}
\eeq
({\em cf.\ e.g.\/} \cite{Chremark} for a proof under the present asymptotic
conditions).  By (\ref{X.0}) we have $p_\mu A^\mu = 0$. Now, by
Prop. \ref{PN.1},  $p_\mu $ is proportional to $ A_\mu$, so if $A^\mu
A_\mu \neq 0$ the result follows.
\hfill\ $\Box$

The main result of this section addresses the case of asymptotically
null Killing vectors. Unfortunately the proof below requires
more asymptotic regularity than one would wish to have. It would be
of some interest to find out whether or not the result below is sharp,
in the sense that decay conditions on three derivatives of the metric
and two derivatives of the extrinsic curvature are
necessary.

\begin{Theorem}\label{TV.1} 
Let $R>0$ and let $(g_{ij},K_{ij})$ be initial data on  
$\Sigma _R={\Bbb R}^3\backslash B(R)$ satisfying 
\begin{eqnarray} &
g_{ij}-\delta_{ij}=O_{3+\lambda}(r^{-\alpha}), \qquad
K_{ij}=O_{2+\lambda}(r^{-1-\alpha}), & \label{V.0}  \\
& J^i = O _{1+\lambda}(r^{-3-\epsilon}), \qquad \rho = O_{1+\lambda}
(r^{-3-\epsilon}), 
\\ 
&\alpha> 1/2, \qquad \epsilon >0, \qquad  0<\lambda<1. & \nonumber
\end{eqnarray}
Let $N$ be a scalar field and $Y^i$  a vector field on $\Sigma_R$ such that
$$ 
N\to_{r\to\infty}A^0,\quad Y^i\to_{r\to\infty}A^i,
\qquad A^\mu A_\mu=0\ , 
$$ 
for some constants $A^\mu\not \equiv 0$. Suppose further that
\begin{eqnarray} &
2 N  K_{ij} + \cL_Y g_{ij} =  O_{3+\lambda}
(r^{-2-\epsilon})\ , & \label{K.1500} 
\\ & 
\tau_{ij} =O_{1+\lambda}
(r^{-3-\epsilon})\ ,  & \label{K.1501} 
\end{eqnarray}
where $\tau_{ij}$ is defined by the equation
\begin{eqnarray}
  & N( \tau_{ij}-\frac{1}{2}g^{k\ell}\tau_{k\ell}g_{ij}) =
N({}^3R_{ij} + K K_{ij} - 2 K_{ik} K^k{}_j)
\qquad\qquad\qquad\qquad \nonumber
\nonumber & \\ &
\qquad \qquad \qquad\qquad\qquad 
- \cL_Y K_{ij} + D_i D_j N  -\frac{\rho}{2}\,N\,g_{ij} \ . &
\label{K.15010}
\end{eqnarray}
Then the ADM four-momentum of $\Sigma _R$ vanishes. 
\end{Theorem}

{\bf Remark}:
There is little doubt that the result is still true with
$\lambda=0$. To prove that one would, however, need to extend the weighted
Sobolev estimates of
\cite{Bartnikmass} to the case $\mbox{\rm dim}M =2$, a task which
lies beyond the scope of this paper.

\proof 
Arguments similar to the proof of Proposition \ref{2PN.1}, Appendix 
\ref{Aproof}, show that
$$ 
N-A^0 = O_{3+\lambda} (r^{-\alpha}), \quad
Y^i-A^i = O_{3+\lambda} (r^{-\alpha})\ .
$$ 
Rescaling $A^\mu$ if necessary we can choose the  
coordinate system so that $A^0=1, A^i=\delta _z^i$. Replacing $\ep$ by  
any number smaller than one if necessary we can assume that $\ep<1$
and $\epsilon\le 2\alpha-1$.  
Taking the trace of eq.\ (\ref{K.1501}) and using the scalar constraint  
equation we find 
$$ 
\Delta_\delta  N+K_{,z}=O_{1+\lambda}(r^{-3-\ep}). 
$$ 
Here, as before, $\Delta_\delta =\partial_x^2+\partial_y^2+\partial_z^2$.  
Let $\varphi$ be as in eq.\ (\ref{PN.1.3.0}), we obtain 
$$ 
\Delta_\delta (N-\varphi _{,z})=O_{1+\lambda}(r^{-3-\ep}), 
$$ 
hence there exists a constant $D$ such that 
 \begin{equation}
N-\varphi _{,z}=1+\frac{D}r+O_{3+\lambda}(r^{-1-\ep}). \label{Neq} 
 \end{equation}
In harmonic coordinates eqs.\ (\ref{(PN.1.0)}), (\ref{(PN.1.6)}),
 (\ref{K.1501}) 
and    (\ref{Neq}) give 
 \begin{eqnarray} &
-\frac{1}2 \Delta_2 g_{ij}=\chi_{ij}+\Psi _{ij}, &\label{V.1} 
\\&
\chi _{ij}=-2m \partial_z 
[\delta _z^j \partial_i\frac{1}r+\delta _z^i \partial_j\frac{1}r] 
+\partial_i\partial_j\frac{D}r, & \label{V.1.0}  
\\&
\Psi _{ij}=O_{1+\lambda}(r^{-3-\ep}). & \label{V.2.0} 
 \end{eqnarray}
Here $\Delta_2=\partial_x^2+\partial_y^2$.  In what follows the indices  
$A,B$, etc.\  take values in the set $\{1,2\}$. Consider the eq.\  
\eq{V.1} with $i=z$, $j=A$.   We have  
 \begin{equation}
\Delta_2 g_{zA}=(8m-2D)\partial_A\partial_z\frac{1}r 
+O(r^{-3-\ep}).\label{V.2.1} 
 \end{equation}
It follows from \cite{ChAFT,Meyers} that for every fixed value of $z$  
the functions $g_{zA}$ have the asymptotic expansion 
 \begin{equation}
g_{zA}=C_{AB}(z)\partial_B \ln\rho +O_{(1)}( \rho ^{-1-\ep}\ln \rho). 
\label{V.2.2} 
 \end{equation}
Here $\rho^2=x^2+y^2$, the functions
 $C_{AB}(z)$ are functions of $z$ only, and we write  
\beq
\label{bracketconvention}
f=O_{(1)}( \rho ^{-\alpha } \ln^\beta \rho)\quad \mbox{\rm if } \quad
|f|+\rho |\partial_A f|\leq C
(1+\rho )^{-\alpha }(1+\ln (1+\rho))^\beta
\eeq
 for some constant $C$ 
which may depend upon $z$. Let  
us define $S(\rho,a )$ to be a circle of radius $\rho $ centered at 
$x=y=0$  
lying in the plane $z=a$. 
 Eq.\  (\ref{V.2.2}) shows that for any fixed value of $z$ the limits 
$$ 
\lim_{\rho \to\infty}\int_{S(\rho ,z)} 
g_{zB}dx^C, \qquad 
\lim_{\rho \to\infty}\int_{S(\rho ,z)} 
x^D\partial_A g_{zB}dx^C 
$$ 
exist. It also follows from our asymptotic conditions on  
$g_{ij}$, eq.\  \eq{V.0},  that these limits are $z$--independent. Set 
 \begin{equation}
\Omega=\lim_{\rho \to\infty}\int_{S(\rho ,z)} 
(x^A\partial_C g_{zA}-g_{zC})dx^C.\label{V.2.3} 
 \end{equation}
For $|z|>R$ by the Stokes  theorem we have  
\begin{eqnarray*} &
\Omega =\int_{{\Bbb R}^2}x^A \Delta_2 g_{Az}=(1)+(2) , & \\ &
(1) = (8m-2D)\int_{{\Bbb R}^2} x^A \partial_z\partial_A\frac{1}r, & \\ &
(2) = \int_{{\Bbb R}^2}x^A\Psi _{Az} \, , &
\end{eqnarray*}
$\Psi _{Az}$ - as in \eq{V.1}. The first integral is easily  
calculated and equals
\begin{equation}
8\pi(4m-D)\sgn z,\label{V.2.4} 
 \end{equation}
where $\sgn z$ denotes the sign of $z$. To estimate the second integral  
it is convenient to split the region of integration into the sets  
$\rho \leq |z|$ and $\rho \geq |z|$. One then  finds 
 \begin{equation}
|(2)|\leq C|z|^{-\ep} 
\mbox{ for }\  |z|> R\ , \label{V.2.5} 
 \end{equation}
with a constant $C$ which does {\em not} depend upon $z$.
Equations \eq{V.2.4}--\eq{V.2.5} are consistent with 
$\partial\Omega/\partial z = 0$ if and only if  
 \begin{equation}
4m=D\ . \label{V.2.6} 
 \end{equation}
Consider now eq.\ 
 \eq{V.1} 
with $i=A,j=B$. Differentiating this  
equation with respect to $z$ one obtains 
 \begin{equation}
\Delta_2 
\frac{\partial g_{AB}}{\partial z}= 
-2D \partial_A \partial_B \partial_Z 
\frac{1}r + 
O(r^{-4-\ep}). \label{V.3}. 
 \end{equation}
By hypothesis we have  
$\frac{\partial g_{ij}}{\partial z}=O(r^{-1-\ep})$, and the estimates  
of \cite {ChAFT} or \cite {Meyers} show that there exist functions  
$D_{ABCD}(z)$ such that for any fixed value of $z$ we have 
 \begin{equation}
\frac{\partial g_{AB}}{\partial z}= 
D_{ABCD} \partial_C\partial_D \ln \rho+ 
O_{(1)}( \rho ^{-2-\ep}\ln \rho).\label{V.4}. 
 \end{equation}
 Let us set 
$$ 
\Omega'=\lim_{\rho \to\infty} \int_{S(\rho ,z)} 
(2x^A x^B \partial_C\partial_z g_{AB}-x^A x^A\partial_C\partial_z 
g_{BB}+  
2 x_C \partial_z g_{AB}-4x^B\partial_z g_{CB}) dx^C. 
$$ 
\eq{V.4} shows that $\Omega'$ is well defined, while \eq{V.0}  
implies that $\Omega'$ is $z$--independent. For $|z|>R$ we again use  
the Stokes theorem to obtain  
$$ 
\Omega'= \int_{{\Bbb R}^2} 
(2 x^A x^B \Delta_2 \partial_z g_{AB}-x^A x^A \Delta_2\partial_z g_{BB}). 
$$ 
A calculation as above leads to 
$$ 
\Omega'=16\pi D \sgn z+O(|z|^{-\ep}),\quad |z|>R. 
$$ 
Hence $D=m=0$ ({\em cf.\ eq.\/} \eq{V.2.6}), which 
together with Proposition \ref{PN.1} establishes our claims. 
\hfill $\Box$

\section{A positive energy theorem}
\label{pets}

In this section we shall prove a
``future--pointing--timelike--or--vanishing--energy--momentum--theorem'', 
under conditions
weaker than previously considered. The main two issues we wish to
address  are 1) the impossibility of a null ADM four--momentum
and 2) a result  which invokes hypotheses concerning  only the fields
$g_{ij}$ and $K_{ij}$.

Let us start with an example of a metric with ``null ADM 
four--momentum''.
Recall that in \cite{AS} Aichelburg and Sexl consider  a sequence of
 Schwarzschild space--times with energy--momentum vector
$(m,0,0,0)$. After applying a ``boost'' transformation to the
Schwarzschild space--time  one obtains an energy--momentum vector
$(\gamma m,\gamma v 
m,0,0)$. Then one takes the limit $v\to 1$ keeping $\gamma m$ equal
to a fixed constant $ p$. The resulting space--time has a
distributional metric and it is not clear if it is asymptotically
flat. Nevertheless, it seems reasonable
to assign to the Aichelburg--Sexl solutions a  null
energy--momentum  vector $(p,p,0,0)$. So, in this sense, there exist
space--times with a null energy--momentum vector.

The Aichelburg--Sexl metrics are plane--fronted waves, and it is of
interest to enquire whether any asymptotically flat plane--fronted
wave metrics exist. Recall that the usual approach in defining
asymptotic flatness is to introduce coordinate systems on $
({\R}^3\setminus B(R))$.  
Consider thus a plane--fronted wave metric on ${\R}\times
({\R}^3\setminus B(R))$,  
\beq
\label{pm}
ds^2=-2du\,dz + \alpha\, dz^2+dx^2 + dy^2\ .
\eeq
As is well known ({\em cf.\ e.g.\/} \cite{Brinkmann,Schimming}), the
metric \eq{pm} is vacuum if and only if  
$\alpha=\alpha(x,y,z)$ with
\beq
\label{pm1}
(\partial^2_x + \partial^2_y)\alpha = 0 \ .
\eeq
Let then $\alpha$ be any solution of \eq{pm1} such that
$\alpha= 1$ for $|z|\ge R$, but $\alpha\not \equiv 1$.
Such solutions are easily found, and for any finite $\ell$ we
can choose $\alpha$ to satisfy 
$$
0\le k\le \ell \qquad |\partial_{A_1}\ldots \partial_{A_k}
(\alpha-1)|
\le C r^{-k-1}\ .
$$ An example is given by the function 
\beq
\label{alphafunction}
\alpha = 1+ \phi(z)C^{A_1\ldots
A_\ell}\partial_{A_1} \ldots \partial_{A_\ell} \ln \rho \ ,
\eeq
 where 
$\phi(z)$ is a smooth compactly supported function and $C^{A_1\ldots
A_\ell}$ is a  totally symmetric tensor with constant coefficients.
We have the following:
\begin{enumerate}
\item
If $\ell =1$ the metric \eq{pm} with $\alpha$ given by
\eq{alphafunction} will not satisfy the fall--off
requirements of the 
positive energy--theorem, {\em cf.\/} Theorem \ref{Tpet} below, because
the $z$ derivatives of the metric do not vanish fast enough as
$r$ tends to infinity. This fall--off of
the metric is not known to be sufficient for a well--defined notion of
ADM mass (compare \cite{Bartnikmass,Chremark,ChErice}). However one can 
calculate  the ADM integral \eq{ADMmass}  in the coordinate system
$(x,y,z)$ as above and find that this integral vanishes.
\item
For all  $\ell \ge 2$ the hypersurfaces $u=\mbox{\rm const}$  will
have a well defined vanishing ADM mass.  This does, however, not
follow from our Theorem \ref{TV.1} unless\footnote{Strictly speaking
we would need to have $\ell \ge 4$ to be able to apply Theorem
\ref{TV.1} as is; {\em cf.\/}, however, the remark following that
Theorem. When we know {\em a priori} that the metric is a plane--fronted 
wave,
 we can
use independent arguments to get rid of the H\"older differentiability
index $\lambda$ in Theorem \ref{TV.1}, no details will be given.}
$\ell \ge 3$. 
\end{enumerate}

 Nevertheless this example shows that non--trivial, vacuum, asymptotically 
flat
plane--fronted waves exist (with $p^\mu = 0$), 
as long as no further global conditions
are imposed.

With those examples in mind, let us briefly recall what is known about
the nonexistence of appropriately regular space--times with null
energy--momentum. In \cite{Witten} an argument was given to support
the expectation that the ADM momentum cannot be null for vacuum
or electrovacuum space--times, the general case being left open. In
\cite{AshtekarHorowitz} this case has been excluded under rather
strong global hypotheses on the space--time and under stringent
asymptotic conditions. 
In \cite{Yip} a proof was given assuming only
hypotheses on the initial data. However, the  proof
there is  rather more
complicated than ours. Moreover the asymptotic conditions of
 \cite{Yip} are more restrictive than ours.

We wish next  to emphasize the
following issue: The statement that the ADM mass $m$ is non-negative
requires only the inequality $\rho\ge\sqrt{J_iJ^i}$, where $\rho$ and
$J^i$ are quantities which can be purely defined in terms of the
fields
$g_{ij}$ and $K_{ij}$, {\em cf.\/} eqs. \eq{EP.1}--\eq{EP.2}
below. Now the published
Witten--type proofs that the vanishing of $m$ implies, loosely
speaking, flatness of 
the resulting space--time, involve the full dominant energy condition
($T_{\mu\nu}X^\mu Y^\nu\ge 0$ for all timelike consistently
time--oriented vectors $X^\mu$ and $Y^\nu$) ({\em cf.\ e.g.\/}
\cite{ParkerTaubes}). Recall that the corresponding 
statement of Schoen and Yau  \cite{SchoenYau} does not
involve\footnote{ Their proof, however, requires 
rather strong asymptotic conditions on the fields. Moreover Schoen and
Yau require the trace of the extrinsic curvature to fall--off at least as
$r^{-3}$. In general this can be justified by applying a ``logarithmic
supertranslation'' in time  to the initial data surface, and requires
the supplementary hypothesis  that the associated 
space--time is large enough. Finally to guarantee that all the
required hypotheses hold on the deformed hypersurface one 
needs again the full dominant energy condition.}   any 
supplementary field $T_{\mu\nu}$.  Similarly both the
proof in \cite{AshtekarHorowitz} and the  proof in
\cite{Yip} which exclude 
a null ADM energy--momentum assume
the full dominant energy condition . A result involving only
conditions on $g_{ij}$ and $K_{ij}$ seems to be much more
satisfactory from a conceptual point of view, and it seems reasonable to 
expect
that the desired conclusion could  be obtained in the Witten--type
setting without imposing conditions on fields other than $g_{ij}$ and 
$K_{ij}$. We show below that this is indeed the case.

Before passing to the statement of our results, in addition to the
papers already quoted let us mention the papers
\cite{LV,Jezierski,KijJezier,KijJezier2,GHHP,ChoquetBruhatlesHouches,Bartnikmass,Reula,ReulaTod,MalecBizon,Penrosetwistorletter,PenroseSorkinWoolgar}
where proofs or arguments relevant to the positive
energy--theorem have been given. The review paper \cite{Horowitz}
contains some further references.

We have the following:

\begin{Theorem}[(Rigid) positive energy theorem]
\label{Tpet}
 Consider a data set
 $(\Sigma,g_{ij},K_{ij})$, with $\Sigma$ of the form
$\Sigma =\Sigma _{{\rm int}} \bigcup^I_{i=1}\Sigma _i
$,
for some 
$I<\infty$.  Here we assume that $\Sigma _{\rm int}$ is compact, and
that each
of the ends $\Sigma _i$ is diffeomorphic to ${\R}^3\setminus B(R_i)$ for
some $R_i>0$, with $B(R_i)$ --- coordinate ball of radius $R_i$. In
each of the ends $\Sigma _i$ the fields $(g,K)$ are assumed to
satisfy the following inequalities 
\begin{equation}
\label{falloff}
| g _{ij}-\delta_{ij}|+|r\partial_k  g _{ij}|+|rK_{ij}|  \le C
r^{-\alpha}\ , 
\end{equation}
for some  constants $C>0$ and $\alpha>1/2$, with 
$r=\sqrt{\sum (x^i)^2}$.
  Suppose moreover that the quantities $\rho$ and $J$ 
\begin{eqnarray}
& 2 \rho := {}^3R + K^2 - K^{ij}K_{ij}\,, &
  \label{EP.1}\\
& J^k := D_l (K^{kl} - K g^{kl}) \,, &
  \label{EP.2}
\end{eqnarray}
are well defined (perhaps in a distributional sense), and satisfy
  \begin{equation}
\sqrt{ g _{ij}J^iJ^j}\le \rho \le C(1+r)^{-3-\epsilon}, \qquad \epsilon >0.
    \label{EP.3}
  \end{equation}
Then the 
 ADM four--momentum $(m,p^i)$ of any of the asymptotic ends of $\Sigma$ 
satisfies $m\ge \sqrt{p_ip^i}$. If $m=0$, then  $\rho\equiv J^i \equiv
0$, and
there exists an isometric embedding $i$ of $\Sigma$ into Minkowski
space--time $({\R}^4,\eta_{\mu\nu})$
such that   $K_{ij}$ represents the extrinsic curvature tensor of
$i(\Sigma)$ in  $(M,\eta_{\mu\nu})$. Moreover $i(\Sigma)$ is an
asymptotically flat 
Cauchy surface in $({\R}^4,\eta_{\mu\nu})$.
\end{Theorem}

\proof Under the conditions here the ADM four--momentum of each of the
asymptotic regions of $\Sigma$ is finite and well defined
\cite{ChErice,Bartnikmass}. As discussed {\em e.g.\/} in
\cite{Chremark}, under the boundary conditions here the Witten
boundary integral reproduces correctly the ADM four--momentum. The
arguments of any of the references \cite{Bartnikmass,Reula,Chremark}
show that one can find solutions to the Witten equation which
asymptote to a constant non--zero spinor in one of the asymptotic
ends, and to zero in all the other ones. Witten's identity
subsequently implies that the ADM momentum of each of the ends is
non--spacelike. 

Suppose that in one of the ends $m$ vanishes. Then for each
$\vec n \in {\bf R}^3$ there exists a spinor field $\lambda_M(\vec n)$
defined on $\Sigma$ satisfying eq. (\ref{A.7}), such that the 
corresponding vector field $Y^j(\vec n)$
defined via eq. (\ref{A.8}), and the scalar field $N(\vec n)$ defined
by eq. (\ref{A.9}), satisfy
 $$ 
Y^j(\vec n) {\ra}_{r \ra \infty} \vec n^j, \qquad
N(\vec n) {\ra}_{r \ra \infty}  |\vec n|_\delta.
$$ 
Here $|\vec n|_\delta$ is the norm of $\vec n$ in the flat metric on
${\bf R}^3$. As shown in Appendix \ref{A}, the fields
$N(\vec n)$ and $Y^i(\vec n)$ satisfy the linear system of equations
({\em cf.\/} eqs.\ (\ref{A.11}) and (\ref{A.11.0}))
\beqa
D_i Y_j + N K_{ij} &=& 0, \label{mo.1} \\
D_i N + K_{ij} Y^j &=& 0. \label{mo.2}
\eeqa
Consider the fields
\beqa
Y_j &=& Y_j((1/2,1/2,0)) - Y_j((-1/2,1/2,0)) - Y_j((1,0,0)),
\label{mo.3} \\
N &=& N((1/2,1/2,0)) - N((-1/2,1/2,0)) - N((1,0,0)). \label{mo.4}
\eeqa
The fields $Y_j$ and $N$ satisfy eqs. (\ref{mo.1})--(\ref{mo.2}) by
linearity of those equations. Moreover we have
\beq
Y^j {\ra}_{r \ra \infty}  0, \qquad
N {\ra}_{r \ra \infty}  1. \label{mo.5}
\eeq
Let $(\wh M,\wh g_{\mu\nu})$ be the Killing development of
$(\Sigma,g_{ij},K_{ij},N,Y_i)$. As discussed in Section \ref{KVSH},
it follows from eqs.\ (\ref{mo.1})--(\ref{mo.2}) that the vector field
$X^\mu \partial_\mu = \partial_u$ is covariantly constant on $\wh M$;
(\ref{mo.5}) implies then
\beq
\wh g_{\mu\nu} X^\mu X^\nu = -1 \quad \Longrightarrow \quad N^2
 - g_{ij} Y^i Y^j = 1. \label{mo.6}
\eeq
By Proposition 3.1 of \cite{ChWald} $\Sigma$ is a Cauchy surface for
$(\wh M, \wh g_{\mu\nu})$. We wish to show that $(\wh M,\wh g_{\mu\nu})$
is geodesically complete. Consider, then, an affinely parametrized
geodesic $x^\mu(s)$, and let $p$ denote the constant of motion
associated with the Killing vector $X^\mu$:
\beq
p = \wh g_{\mu\nu} \dot x^\mu X^\nu = - \dot u + Y_i \dot x^i.
\label{mo.7}
\eeq
Here eqs. (\ref{K.10}) and (\ref{mo.6}) have been taken into account;
a dot over a quantity means differentiation with respect to $s$. 
Since $s$ is an
affine parameter we have, with $\ve = 0,\pm 1$,
\beq
- \dot u^2 + 2 Y_i \dot x^i \dot u + g_{ij} \dot x^i \dot x^j = \ve.
\label{mo.8}
\eeq
Eqs. (\ref{mo.7}) and (\ref{mo.8}) give
\beq
(g_{ij} + Y_i Y_j) \dot x^i \dot x^j = \ve + p^2.
\label{mo.9}
\eeq
(\ref{mo.9}) and (\ref{mo.8}) imply that there exists a function $C(p)$
such that
\beq
|\dot x|_g + |\dot u| \leq C(p). \label{mo.10}
\eeq
Choose $p \in {\bf R}$ and consider the set $\Omega_p$ of
maximally extended affinely parametrized geodesics with that value of
$p$, with $x^\mu(0) \in \Sigma$. We can without loss of generality assume
that $\alpha < 1$; an analysis of eqs.\ \eq{mo.1}--\eq{mo.2}
along the lines of Appendix \ref{Aproof} shows that $\wh g_{\mu\nu} - \eta 
_{\mu\nu}= O_1(r^{-\alpha})$. By asymptotic flatness of
$\wh g_{\mu\nu}$ (cf. Proposition \ref{2PN.1}) and the interior compactness
condition on $\Sigma$ there exists $\delta > 0$ such that all geodesics
in $\Omega_p$ are defined for $s \in (-\delta,\delta)$. Eq. (\ref{mo.10})
shows that in that affine time the value of $|u|$ can change at most by
$C(p)\delta$, similarly for the value of $r(s) \equiv (x^2(s) + 
y^2(s) + z^2(s))^{1/2}$ in the asymptotic regions. One can now invoke
the fact that the $u$-translations are isometries to conclude that
all geodesics in $\Omega_p$ are complete, and the result follows.

Let us show now that $(\wh M,\wh g)$ is flat. Let
$Y^i_{(k)} = Y^i(\vec e_k)$, where $Y^i(\vec n)$ is as at the beginning
of this proof and where the $\vec e_k$'s, $k = 1,2,3$, form an orthonormal
basis of ${\bf R}^3$. Let $N_{(k)} = N(\vec e_k)$ be the corresponding
lapse functions. On $\wh M$ define the fields $X^\mu_{(k)}$ by the eq.
\beq
X^\mu_{(k)} \partial_\mu = \wh N_{(k)} n^\mu \partial_\mu + 
\wh Y^i_{(k)} \partial_k, \label{PET.1} 
\eeq
$$
\wh Y^i_{(k)}(u,x^i) = Y^i_{(k)} (x^i), \qquad
\wh N_{(k)}(u,x^i) = N_{(k)}(x^i).
$$
Here $n^\mu$ is the field of unit normals to the slices
$\{ u = \mbox{const}\}$. By eqs. (\ref{A.11}) and (\ref{A.11.0}) we
have
\beq
\wh \nabla_j X^\mu_{(k)} = 0 . \label{PET.2}
\eeq
By construction of $(\wh M,\wh g_{\mu\nu})$ it also holds that
\beq
\wh \nabla_\mu X^\nu = \wh \Gamma^\nu_{\mu\lambda} X^\lambda = 
\wh \Gamma^\nu_{\mu u} = 0. \label{PET.3}
\eeq
As the components of $X^\mu_{(i)}$ are $u$-independent by
(\ref{PET.1}), eq. (\ref{PET.3}) gives
\beq
\wh \nabla_u X^\mu_{(k)} = \partial_u X^\mu_{(k)} + 
\wh \Gamma^\mu_{\lambda u} X^\lambda_{(k)} 
= 0.
\eeq
Consequently
\beq
\wh \nabla_\mu X^\nu_{(k)} = 0 . \label{PET.4}
\eeq
Differentiating (\ref{PET.4}) one obtains
\beq
\wh R_{\mu\nu\rho\sigma} X^\sigma_{(i)} = 0. \label{PET.5}
\eeq
As the vector fields $X^\sigma_{(i)}$ are everywhere null and linearly
independent, standard algebra gives
\beq
\wh R_{\mu\nu\rho\sigma} \equiv 0.
\eeq

Consider, next, the universal covering space $\wt \Sigma$ of $\Sigma$
with fields $(\wt g_{ij},\wt K_{ij},\wt Y_i,\wt N)$ obtained by
pull-back. Let $(\bar M,\bar g_{\mu\nu})$ be the Killing development
of $(\wt \Sigma, \wt g_{ij}, \wt K_{ij}, \wt Y_j,\wt N)$. Clearly
$\bar M$ is the universal covering space of $\wh M$ with $\bar g_{\mu\nu}$ 
being the pull-back of $\wh g_{\mu\nu}$. It is easily seen that
$(\bar M,\bar g)$ inherits from $(\wh M,\wh g)$ the following
properties:
\begin{enumerate}
\item $(\bar M,\bar g_{\mu\nu})$ is globally hyperbolic with Cauchy
surface $\wt \Sigma$.
\item $(\bar M,\bar g_{\mu\nu})$ is geodesically complete.
\item $(\bar M,\bar g_{\mu\nu})$ is flat.
\end{enumerate}
As $\bar M$ is simply connected, it follows {\em e.g.\/} from 
\cite[Theorem 2.4.9]{Wolf}
 that $(\bar M,\bar g_{\mu\nu})$ is the Minkowski space-time $({\bf 
R}^4,\eta_{\mu\nu})$. 
As $\wt \Sigma$ is a Cauchy surface
for $\bar M$, it is necessarily a graph over a spacelike plane
$t = 0$ in $({{\Bbb R}}^4,\eta_{\mu\nu})$. In particular $\wt \Sigma$
has only one asymptotically flat end (compare also \cite[Lemma 2]{Chmass}).
If $\Sigma$ had been non-simply connected, then $\wt \Sigma$ would 
have had more than one asymptotic end. It follows that
$\Sigma = \wt \Sigma$, $\wh M = {{\Bbb R}}^4$ and our claims follow. 
\hfill\ $\Box$

To exclude the case of a null ADM four--momentum we need to
assume some further asymptotic regularity conditions:

\begin{Theorem}
\label{Tpet2}
Under the hypotheses of Theorem \ref{Tpet}, suppose moreover that in 
some of the asymptotic ends it holds that
\begin{eqnarray} &
g_{ij}-\delta_{ij}=O_{3+\lambda}(r^{-\alpha}), \qquad
K_{ij}=O_{2+\lambda}(r^{-1-\alpha}), & \label{V.00}  \\
& \rho=O_{1+\lambda}(r^{-3-\ep}), & \label{V.01}
\end{eqnarray}
with some $0<\lambda<1$. Then the ADM four--momentum of that
end cannot be null.
\end{Theorem}

{\bf Remark:} It can be shown by rather different techniques that the 
result is still true with $\lambda=0$, we shall however not discuss 
that here.
\proof
Consider an asymptotic end $\Sigma_1$ in which
eqs. \eq{V.00}--\eq{V.01} hold and which has a null ADM
four--momentum $p^\mu$.
As discussed in the proof of Theorem \ref{Tpet} and in Appendix
\ref{A}, the hypotheses of Proposition \ref{PA.1} and 
Corollary \ref{CA.1} are satisfied. 
We can thus apply
Theorem \ref{TV.1} to conclude that the ADM four--momentum
of the end under consideration vanishes, and the result follows
from Theorem \ref{Tpet}.
\hfill $\Box$

Let us close this section by proving Theorem \ref{T1}: By the arguments 
given above $\rho$ and $J^i$ vanish on $\Sigma$. It follows
from a result of Hawking and Ellis \cite[Chapter 4, Section 4.3]{HE}
that $(M,g_{\mu\nu})$ must be flat. By uniqueness of the maximal globally 
hyperbolic vacuum developments it follows that the Killing development
constructed in the proof of Theorem \ref{Tpet2} ({\em cf.\/} Appendix 
\ref{A})
coincides with the maximal globally hyperbolic development
of $(\Sigma, g_{ij}, K_{ij})$, and Theorem
\ref{T1} follows.

  {\bf Acknowledgements}
Part of the work on this paper was done when both authors were
visiting the Max Planck Institut f\"ur Astrophysik in Garching; they
are grateful to J\"urgen Ehlers and to the members of the Garching
relativity group for hospitality. P.T.C.\ acknowledges the hospitality
of the E. Schr\"odinger Institute and of the Vienna
relativity group during part of work on this paper.



\appendix
\section{Definitions and conventions}
\label{definitions}
\label{asymptotic}

We say that $(M,g_{\mu\nu})$ is a $C^k$ spacetime if  $M$ is a
paracompact, connected, Hausdorff, orientable manifold of $C^k$
differentiability class, with a $C^{k-1}$ Lorentzian metric.
We use the signature $(-,+,+,+)$.

Consider a function $f$ defined on
$
\Sigma_R\equiv {\R}^3\setminus B(R)
$,
where $B(R)$ is a closed ball of radius $R>0$.
 We shall  write  $f = O_k(r^\beta)$ if there exists a
constant $C$ such that we have
$$
0 \leq i \leq k \qquad |\partial^i f| \leq C r^{\beta - i}.
$$
For $\sigma\in(0,1)$ we shall write
  $f = O_{k+\sigma}(r^\beta)$ if  $f = O_k(r^\beta)$ and if there exists a
constant $C$ such that we have
$$
|y-x|\le r(x)/2 \quad \Rightarrow \quad
 |\partial^k f(x) - \partial^k f(y)| \leq C |x-y|^{\sigma}
r^{\beta - k-\sigma}.
$$
Let us note that $f = O_{k+1}(r^\beta)$ implies $f =
O_{k+\sigma}(r^\beta)$
for all $\sigma\in(0,1)$, so that  the reader unfamiliar with H\"older type
spaces might wish to simply replace, in the hypotheses of our theorems,
 the $k+\sigma$ by $k+1$ wherever
convenient.

\section{Appendix}\label{A}

In this appendix we prove a differential geometric proposition on
initial-data sets $(\Sigma,g_{ij},K_{ij})$ having a nowhere vanishing 
spinor
field which is covariantly constant on $\Sigma$ with respect to the
``Sen-connection'' \cite{Sen} ({\em cf.\/} eq.\  \eq{A.7} below).
 This result forms the local input of the rigidity part of
the positive-mass theorem. Similar results in the literature we are
aware of implicitly or explicitly use Cauchy developments
$(M,g_{\mu\nu},\phi^A)$ of $(\Sigma,g_{ij},K_{ij}, \psi^A)$, for some
fields $\phi^A$ with Cauchy data $\psi^A$, with energy--momentum tensor
$T_{\mu\nu}$
satisfying the {\em full\/} dominant energy condition ({\em cf.\/} the
discussion at the beginning of Section \ref{pets}).
For our results below neither the existence of such a Cauchy
 evolution\footnote{In the case of a
``bad'' matter model --- such as, {\em e.g.}, dust as a source for the 
Einstein
equations -- an evolution is not known to
exist. Similarly even for ``good'' models, such as vacuum Einstein 
equations, the differentiability hypotheses on the initial data in Theorem 
\ref{Tpet} and
\ref{Tpet2} are not known to guarantee existence of a Cauchy 
development.} nor in fact the DEC for the given triple 
$(\Sigma,g_{ij},K_{ij})$ ({\em i.e.},
$\sqrt{g_{ij} J^i J^j} \leq \rho$) is required.

To motivate our three--dimensional discussion, we shall as before start
with the four--dimensional picture. 
Consider thus a spacetime $(M,g_{\mu\nu})$ with $g_{\mu\nu}$ in
$C^2$ and a nowhere zero $C^2$ spinor field $\lambda_M$ on $M$ satisfying
\beq
\label{A.1}
\nabla_\mu \lambda_N = 0 \Longleftrightarrow \nabla_{MM'} \lambda_N = 0,
\eeq
{\em i.e.}, $\lambda_M$ is covariantly constant. 
 We use capital letters in the
second half of the alphabet to denote spinor indices. Since the
considerations in this appendix are purely local, there is no question
of existence of a 
spinor structure. The spinorial Ricci identities ({\em cf.\/}
\cite[Vol.\ 1, pp. 242--244]{PenroseRindler}) immediately imply that
the Ricci scalar  
$R_\mu{}^\mu$ of $g_{\mu\nu}$ is zero, and that the spinor equivalent of
$S_{\mu\nu} := R_{\mu\nu} - \frac{1}{4} g_{\mu\nu} R_\lambda{}^\lambda$,
namely the hermitian spinor $\phi_{MNM'N'}$, satisfies
\beq\label{A.2}
\lambda^M \phi_{MNM'N'} = 0
\eeq
$$
\Longrightarrow \ve_{MN} \lambda^P \phi_{PRR'(M'} \bar \lambda_{N')}
+ \lambda_{(M} \phi_{N)RP'R'} \bar \lambda^{P'} \ve_{M'N'} = 0\ .
$$
This last equation, in tensor terms, says that
\beq\label{A.3}
X_{[\mu} S_{\nu]\lambda} = 0,
\eeq
where $X^\mu$ is the null vector corresponding to 
$\lambda_M \bar \lambda_{M'}$. Consequently
\beq\label{A.4}
R_{\mu\nu} = \sigma X_\mu X_\nu\ ,
\eeq
for some function $\sigma$ on $M$. By eq.\  \eq{A.1}, $X^\mu$ is 
covariantly
constant, {\em i.e.},
\beq\label{A.5}
\nabla_\mu X_\nu = 0, \qquad \mbox{ with} \quad g_{\mu\nu} X^\mu X^\nu = 0.
\eeq
According to one of several equivalent definitions ({\em cf.\ e.g.\/}
\cite{Schimming}), eqs.\ \eq{A.4}--\eq{A.5} imply that $(M,g_{\mu\nu})$ is
a pp--spacetime. 
We have recovered the well-known fact ({\em cf.\ e.g.\/} 
\cite{Tod,Taub,Schimming}) that a spacetime admitting a covariantly
constant spinor describes a pp--wave.

Let, next, $\Sigma$ be a spacelike hypersurface of $(M,g_{\mu\nu})$
with unit-normal $n_\mu$. With $n_{MM'}$ being the spinor equivalent
of $n_\mu$, eq.\  \eq{A.1} implies that
\beq\label{A.6}
n_{(M}{}^{M'} \nabla_{N)M'} \lambda_P = 0 \Longleftrightarrow
n_{[\mu} \nabla_{\nu]} \lambda_P = 0.
\eeq
eq.\  \eq{A.6} contains only derivatives tangential to $\Sigma$. 
When
$\lambda_M$ is interpreted as a $SU(2)$-spinor on $(\Sigma,g_{ij},K_{ij})$,
\eq{A.6} can be written as (we use the conventions of Appendix A of
\cite{Ashtekar}), 
\beq\label{A.7}
D_{MN} \lambda_P + \frac{i}{\sqrt{2}} K_{MNPQ} \lambda^Q = 0,
\eeq
where $K_{MNPQ}$ is the $SU(2)$-spinor version of $K_{ij}$ and 
$D_{MN}$ the covariant derivative on $\Sigma$ associated with $g_{ij}$.

Let us turn to the three--dimensional formulation of the problem.
Suppose  that we are given $(\Sigma,g_{ij},K_{ij})$ with $g_{ij}$ in $C^k$,
for some $k\ge 1$,  $K_{ij}$
in $C^{k-1}$ and a $C^k$-spinor $\lambda_M$ satisfying
eq.\  \eq{A.7}.  We want to embed $\Sigma$ into some Lorentz manifold
$(M,g_{\mu\nu})$ in which $\lambda_M$ extends to a spinor field
obeying eq.\  \eq{A.1}.

Denote by $M_i$ the complex--valued null vector field on $\Sigma$ 
associated with
$\lambda_M \lambda_N$ and define a real vector $Y_i$ by
\beq\label{A.8}
Y_i = \frac{i}{\sqrt{g_{jk} M^j \bar M^k}} \; \ve_i{}^{jk} M_j 
\bar M_k \ ,
\eeq
and a real positive scalar $N$ by
\beq\label{A.9}
N = \sqrt{g_{ij} M^i \bar M^j} = \sqrt{g_{ij} Y^i Y^j}.
\eeq
By, {\em e.g.}, \cite[Lemma 4.3]{ParkerTaubes}
$\lambda_N$ is nowhere zero, hence $N$ is nowhere vanishing.
{}From \eq{A.7}, $M_i$ satisfies 
\beq\label{A.10}
D_i M_j = - i \ve^{\ell m}{}_j K_{i\ell} M_m ,
\eeq
which, after some calculation, implies
\beq\label{A.11}
D_i Y_j + N K_{ij} = 0.
\eeq
We also note, for use in the body of the paper, the equation
\beq\label{A.11.0}
D_i N + K_{ij} Y^j = 0,
\eeq
which follows from (\ref{A.9}) and (\ref{A.11}).
Now define $(\hat M,\hat g_{\mu\nu})$ to be the Killing development
$({{\Bbb R}} \times \Sigma, \wh g_{\mu\nu})$ of $(\Sigma,g_{ij},K_{ij})$ 
based on
$(N,Y^i)$, {\em i.e.},
\beq\label{A.12}
\wh g_{\mu\nu} dx^\mu dx^\nu = - N^2(x^\ell) du^2 + g_{ij}(x^\ell)
[dx^i + Y^i(x^\ell)du][dx^j + Y^j(x^\ell)du].
\eeq
This, as shown in Section \ref{KVSH}, has $X = \partial/\partial u$ as a
covariantly constant null vector, the induced metric on $u = 0$
coincides with $g_{ij}$, and the extrinsic curvature is $K_{ij}$.
The field of 
unit normals $n_\mu$ to the hypersurfaces $\{u = \mbox{ \rm const}\}$ is
Lie derived by this Killing, vector field,
\beq\label{A.13}
\cL_X n_\mu = 0\ ,
\eeq
which can be seen as follows: By construction $X(u) = 1$. Since Lie 
derivation and 
exterior differentiation commute, we have that $\cL_X du = 0$. By the
Killing property of $X$, $\cL_X (du,du)$ is also zero, and
eq.\  \eq{A.13} follows. But, by the covariant constancy of $X$, {\em 
i.e.},
\beq\label{A.14}
\wh \nabla_\mu X_\nu = 0,
\eeq
this implies that
\beq\label{A.15}
 X^\nu \wh \nabla_\nu n_\mu = 0.
\eeq
Now extend $\lambda_M$ off $u = 0$ to a spinor field $\wh \lambda_M$
on $(M,\wh g_{\mu\nu})$ by requiring
\beq\label{A.16}
X^\mu \wh \nabla_\mu \wh \lambda_M = 0.
\eeq
Consider the expression
\beq\label{A.17}
U_{MNP} = n_{(M}{}^{M'} \wh \nabla_{N)M'} \wh \lambda_P.
\eeq
By eqs.\ \eq{A.6}--\eq{A.7}, $U_{MNP}$ vanishes for $u = 0$. Now compute
\beq \label{A.18}
X^\mu \wh \nabla_\mu U_{MNP} =
n_{(M}{}^{M'} X^\mu \wh \nabla_{|\mu|} \wh \nabla_{N)M'} 
\wh \lambda_P,
\eeq
where we have used \eq{A.15}. Since $X$ is covariantly constant,
$X^\mu \wh \nabla_\mu$ commutes with covariant differentiation.
Applying this on the r.h. side of \eq{A.18} and using \eq{A.15}, we infer
\beq\label{A.19}
X^\mu \wh \nabla_\mu U_{MNP} = 0.
\eeq
Thus
\beq \label{A.20}
n_{(M}{}^{M'} \wh \nabla_{N)M'}  \wh \lambda_P = 0 \qquad
\Longleftrightarrow \qquad
n_{[\mu} \wh \nabla_{\nu]} \wh \lambda_P = 0 . 
\eeq
By \eq{A.16} we also have that
\beq \label{A.21}
(N n^\mu \wh \nabla_\mu + Y^i \wh \nabla_i) \wh \lambda_M = 0.
\eeq
Due to \eq{A.20} the second term in \eq{A.21} is zero.
As $N$ is nowhere vanishing we obtain
\beq \label{A.22}
n^\mu \wh \nabla_\mu \wh \lambda_P = 0.
\eeq
Since $n^\mu$ is timelike and again using \eq{A.20} we get
\beq \label{A.23}
\wh \nabla_\mu \wh \lambda_P = 0,
\eeq
as promised. Combining the above calculation\footnote{Strictly speaking 
the above calculations require $k\ge 2$. One can use a slightly different 
argument to show that Proposition \ref{PA.1} is correct as stated.}
 with eq.\  \eq{A.4}, we obtain the following:

\begin{Proposition}
\label{PA.1} Let $k\ge 1$ and let $(\Sigma,g_{ij},K_{ij})$, $g_{ij} \in 
C^k$, 
$K_{ij} \in C^{k-1}$
be such that there exists a  $C^k$ spinor field satisfying
eq.\  \eq{A.7}. Then there exists a nowhere zero vector field $Y_i$ in
$C^k$ such that 
\beq
 D_i Y_j + N K_{ij} = 0\ ,
 \label{A.25}\eeq
where $N := \sqrt{g_{ij} Y^i Y^j}$.
If moreover $k\ge 2$, then the fields 
$(\rho,J_i,\tau_{k\ell})$ defined in eqs.\ \eq{K.13}--\eq{K.15} satisfy
\beq \label{A.24}
N J_i = - \rho Y_i\ , \qquad
N^2 \tau_{ij} = \rho Y_i Y_j\ . 
\eeq
\end{Proposition}

In the case where the 
ADM 4-momentum $p^\mu$ is null, the Witten argument gives rise to a
spinor field on $\Sigma$ obeying eq.\  \eq{A.7}
({\em cf.\/} the discussion and the references in the proof of Theorem
\ref{Tpet}). Proposition 
\ref{PA.1} and an analysis of eqs.\ \eq{A.11}--\eq{A.11.0}  
similar to that of  Appendix \ref{Aproof} lead to the following:

\begin{Corollary}
\label{CA.1} Let $(\Sigma,g_{ij},K_{ij})$ satisfy the hypotheses of
Theorem \ref{Tpet} and let $p^\mu$ be null. Then there exists a nowhere
zero $C^1$-field $Y_i$ with $Y^i - A^i = O_1(r^{-\alpha})$ for some
constants $A^i$, so that eq.\ \eq{A.25} holds. If moreover the hypotheses 
of Theorem \ref{Tpet2} are satisfied, then $Y^i - A^i = O_3(r^{-\alpha})$, 
and \eq{A.24} holds.
\end{Corollary}

\section{Proof of Proposition 
\protect\ref{2PN.1}} 
\label{Aproof}
Eq. (\ref{K.4}) gives the equation
\beq \label{L.1}
D_i D_j Y_k = R_{mijk} Y^m + D_k(N K_{ij}) - D_i(N K_{jk}) -
D_j(N K_{ki}).
\eeq
Here $R_{mijk} $ is the curvature tensor of the metric $g_{ij}$.
Consider the system of equations
\beqa
\frac{\partial N}{\partial r} &=& \frac{x^i}{r} \partial_i N  \ , 
\label{L.2} \\
\partial_r Y_i &=& \frac{x^j}{r} (D_j Y_i + \Gamma^k_{ij} Y_k) \ 
,\label{L.3} \\
\partial_r D_i N &=& \frac{x^k}{r} (D_k D_i N +
\Gamma^j_{ki} D_j N) \ , \label{L.4} \\
\partial_r D_i Y_j &=& \frac{x^k}{r} (D_k D_i Y_j +
\Gamma^\ell_{ki} D_\ell Y_j + \Gamma^\ell _{kj} D_i Y_\ell) \ . \label{L.5}
\eeqa
Here we are implicitly assuming that in \eq{L.4} and \eq{L.5} the terms
$D_k D_i N$ and $D_k D_i Y_j$ have been eliminated
using \eq{K.15} and \eq{L.1}. Set $f = (f^A) = (N, r D_i N, Y_j, r D_i 
Y_j)$,
$g = \sum_A f^A f^A$. We have
\beq \label{L.5.1}
\left| \frac{\partial g}{\partial r} \right| \leq \frac{C g}{r},
\eeq
and by $r$-integration one finds
\beq\label{L.6}
|f| \leq C(1 + r^\beta)
\eeq
for some constants $C,\beta$. Suppose that $\beta > 2$, using \eq{L.6} and
\eq{L.2}--\eq{L.5} one finds by $r$-integration $|f| \leq 
C(1 + r^{\beta - \alpha})$, so that \eq{L.6} has been improved by 
$\alpha$. 
Iterating this process one obtains \eq{D.4.0} and \eq{D.4.1}, {\em cf.\/} 
also \cite[Appendix A, Lemma]{Chmass}.
Suppose finally that $A^\mu = \Lambda_{\mu\nu} = 0$. Iterating further
one finds
\beq\label{L.7}
|f| \leq C r^{-\sigma} \qquad \mbox{for any } \sigma > 0.
\eeq
Note that if $g(r_0) = 0$, at some $r_0$, then by \eq{L.5.1} we will have
$g \equiv 0$. Suppose thus that for all $r$ there holds $g(r) \neq 0$. For
$r_1 \geq r_0$ we then have by \eq{L.5.1} 
$$
\frac{\partial g}{\partial r} \geq -\frac{C g}{r} \Ra \ln (g(r_1)
r_1^C) \geq \ln (g(r_0) r_0^C).
$$
Passing with $r_1$ to infinity from \eq{L.7} we obtain $g(r_0) = 0$, which
gives a contradiction, and the result follows. \hfill\ $\Box$

\end{document}